\documentclass[twocolumn,a4paper,10pt,showkeys,aps,pra,longbibliography,nofootinbib,superscriptaddress]{revtex4-1}

\usepackage[utf8x]{inputenc}
\usepackage[UKenglish]{babel}
\usepackage[UKenglish]{isodate}
\usepackage{graphicx}
\usepackage{exscale}
\usepackage{color}
\usepackage{amsfonts}
\usepackage{amssymb}
\usepackage{mathptmx}
\usepackage{textcomp}

\begin{document}

\title{Influence of ionisation zone motion in high power impulse magnetron sputtering on angular ion flux and NbO$_x$ film growth}

\author{Robert Franz}
\email[Corresponding author: ]{robert.franz@unileoben.ac.at}
\affiliation{Lawrence Berkeley National Laboratory, 1 Cyclotron Road, Berkeley, California 94720, USA}
\affiliation{Montanuniversit\"{a}t Leoben, Franz-Josef-Strasse 18, 8700 Leoben, Austria}
\author{César Clavero}
\affiliation{Lawrence Berkeley National Laboratory, 1 Cyclotron Road, Berkeley, California 94720, USA}
\author{Jonathan Kolbeck}
\affiliation{Lawrence Berkeley National Laboratory, 1 Cyclotron Road, Berkeley, California 94720, USA}
\author{André Anders}
\affiliation{Lawrence Berkeley National Laboratory, 1 Cyclotron Road, Berkeley, California 94720, USA}

\begin{abstract}
\noindent 
The ion energies and fluxes in the high power impulse magnetron sputtering plasma from a Nb target were analysed angularly resolved along the tangential direction of the racetrack. A reactive oxygen-containing atmosphere was used as such discharge conditions are typically employed for the synthesis of thin films. Asymmetries in the flux distribution of the recorded ions as well as their energies and charge states were noticed when varying the angle between mass-energy analyser and target surface. More positively charged ions with higher count rates in the medium energy range of their distributions were detected in $+\mathbf{E}\times \mathbf{B}$ than in $-\mathbf{E}\times \mathbf{B}$ direction, thus confirming the notion that ionisation zones (also known as spokes or plasma bunches) are associated with moving potential humps. The motion of the recorded negatively charged high-energy oxygen ions was unaffected. NbO$_x$ thin films at different angles and positions were synthesised and analysed as to their structure and properties in order to correlate the observed plasma properties to the film growth conditions. The chemical composition and the film thickness varied with changing deposition angle, where the latter, similar to the ion fluxes, was higher in $+\mathbf{E}\times \mathbf{B}$ than in $-\mathbf{E}\times \mathbf{B}$ direction.
\end{abstract}


\cleanlookdateon
\date{\today}

\maketitle

\section{Introduction}

More than a decade ago, high power impulse magnetron sputtering (HiPIMS) was introduced as a variation of conventional dc magnetron sputtering (dcMS) \cite{Kouznetsov1999}. HiPIMS is typically characterised by high discharge peak currents and a high degree of ionisation of the sputtered species enabling energetic growth conditions with enhanced control possibilities during film deposition. The method has been reviewed several times and further details can be found in the following reviews: Helmersson \textit{et al.}~\cite{Helmersson2006}, Sarakinos \textit{et al.}~\cite{Sarakinos2010}, Gudmundsson \textit{et al.}~\cite{Gudmundsson2012} and Britun \textit{et al.}~\cite{Britun2014}.

In recent years, optical imaging using high speed cameras revealed the presence of plasma instabilities along the racetrack of the target, so-called plasma bunches, ionisation zones, spokes or emission structures \cite{Kozyrev2011,Anders2012a,Ehiasarian2012,Winter2013}. These zones of intense plasma follow the $\mathbf{E}\times \mathbf{B}$ drift (other drifts in the same direction are included) of the electrons with a velocity of about 1/10 of the electrons' drift velocity and their motion seems to be a consequence of the ``evacuation'' of ions from their location of production (ionisation) \cite{Anders2012d}. The number and the shape of the zones depend on the discharge current \cite{Yang2015,Poolcharuansin2015}. Typically, the zones reveal an arrow-like shape with the tip pointing in the direction of the motion, but at higher currents it changes to a rather globular appearance which appears to be associated to a change of the ionization zones' plasma and potential distribution in the direction perpendicular to the target, as fast imaging techniques suggest \cite{Ni2012}. 

A phenomenological model of the ionisation zones was introduced by Gallian \textit{et al.}~to qualitatively described the observed effect \cite{Gallian2013a}. In order to explain the different shapes, triangular (arrow-like) or diffuse (globular), Hecimovic \textit{et al.}~presented a model that is based on the localised generation of secondary electrons \cite{Hecimovic2014}. The configuration of the electric field along the target's racetrack was first discussed by Brenning \textit{et al.}~\cite{Brenning2013} and later modified by Anders \textit{et al.}~\cite{Anders2013,Anders2014a}. In the latter publication it is argued that for a closed racetrack, the potential needs to be reproduced when returning to the same location. This implies the presence of a double layer enclosing the ionisation zone with negative space charges at the outside and positive space charges in the centre. The resulting potential hump has a maximum, most likely at the place of maximum plasma intensity, and appears to be asymmetric according to the recorded optical images of the ionisation zones. As a consequence of such an electric field configuration, the local $\mathbf{E}\times \mathbf{B}$ drift causes an acceleration of electrons away from the target surface \cite{Anders2014a} which was noticed as the appearance of plasma flares.

Such plasma flares extending away from the target surface into the space of the vacuum chamber were recorded by Ni \textit{et al.}~and it was shown that they are associated to moving ionisation zones \cite{Ni2012}. Spectroscopic imaging with short exposure times revealed that the light emission from excited target atoms and ions (an Al target was used for this study) stems from a region in the vicinity of the target surface, whereas the light emission from excited Ar background gas atoms and ions was more distributed including areas distant from the target \cite{Andersson2013}. Even though the highest concentration of ions from the target material appears close to the target surface within the ionisation zones, the angular dependence of their flux from the racetrack has been recognised to influence the film growth conditions when using HiPIMS discharges to synthesise thin film materials.

When depositing Ti and Cr films on substrates placed on the side of the magnetron and perpendicular to the target surface normal (corresponding to an angle of 90\textdegree~in the current work, but shifted away from target surface), Lundin \textit{et al.}~achieved normalised deposition rates of 0.8 and 0.5 for Ti and Cr, respectively \cite{Lundin2008}. The reference position for the normalisation was opposite to the target surface (0\textdegree~in this work). Compared to dcMS, the HiPIMS deposition rates on the side of the magnetron was higher by 10\% in the case of Ti and by 25\% in the case of Cr. By recording the ion energy distribution functions (IEDFs) at places similar to the substrate positions, the authors also noticed an asymmetry in the ion energies depending on the direction with respect to the racetrack. This asymmetry was confirmed by Poolcharuansin \textit{et al.}~applying a retarded field analyser to measure the ion energies spatially and angularly resolved in planes parallel to the target surface \cite{Poolcharuansin2012}. The obtained mean ion energies from a Ti target had maxima in the direction of the $\mathbf{E}\times \mathbf{B}$ drift and, hence, the potential hump motion direction ($+\mathbf{E}\times \mathbf{B}$ in this work). 

In a previous work, we studied the time- and energy-resolved characteristics of HiPIMS and dcMS discharges from a Nb target in planes parallel to the target surface \cite{Panjan2014}. Integration of the IEDFs revealed that a higher number of Nb$^+$ and even more so of Nb$^{2+}$ ions are emitted in the direction of the electrons' $\mathbf{E}\times \mathbf{B}$ drift. This directionality is a result of the ionisation zone and associated potential hump motion as the doubly charged Nb ions gained roughly double the energy, i.e.~the energy contribution in the IEDFs originating from the acceleration in the potential hump. These results were observed in highly ionised HiPIMS discharges, but to some extend also in conventional dcMS discharges \cite{Panjan2014} although other driving factors play a role which even lead to a reversal of the motion direction of ionization zones \cite{Yang2014}. However, since ions from the target material significantly contribute to the film growth in HiPIMS, we decided to study the angular resolved thin film growth conditions in HiPIMS with the aim of relating them to the newly discovered plasma characteristics, i.e.~the moving ionisation zones and associated potential humps.

\section{Experimental details}

\begin{figure}
 \centering
 \includegraphics[width=8cm]{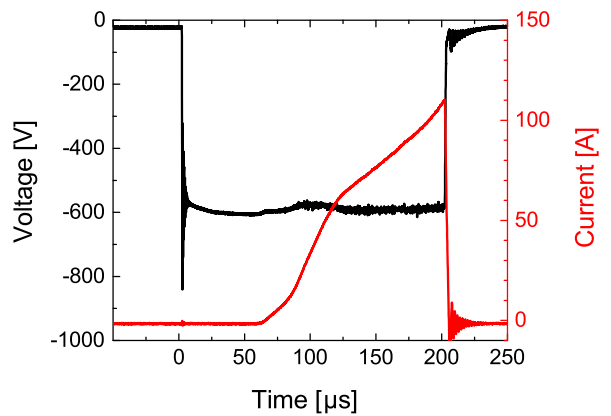}
 \caption{Typical discharge voltage and current signals of the HiPIMS pulse used within this work.}
 \label{fig:voltage-current}
\end{figure}

The measurements of the plasma properties and the deposition of the NbO$_x$ thin films were done in a vacuum chamber that was evacuated to a typical background pressure of $5\cdot 10^{-4}$ Pa. The total pressure during the experiments was set to 0.25 Pa with Ar and O$_2$ flows of 80 and 20 sccm, respectively. The Nb target with a diameter of 7.6 cm (3 inch) was mounted on a water-cooled unbalanced magnetron with a grounded surrounding anode ring that was positioned flush with the target surface. A commercially available pulse generator Melec SIPP 2000 was used to create 200 \textmu s pulses with a repetition rate of 100 Hz. The pulse generator was charged with an Pinnacle DC power supply by Advanced Energy to a charging voltage of $\sim$600 V at an average power of $\sim$500 W. As shown in Fig.~\ref{fig:voltage-current}, the peak current during the HiPIMS pulse was typically 110 A. With the used O$_2$ partial pressure, the HiPIMS discharge was operated in ``poisoned'' mode. The discharge voltage and current characteristics are similar to the ones reported by Leroy \textit{et al.}~\cite{Leroy2011} and Aiempanakit \textit{et al.}~\cite{Aiempanakit2013} who studied reactive HiPIMS plasmas from Ti and Al targets.

The IEDFs were measured with a HIDEN EQP 300, a combined ion energy and mass analyser, pointing at the racetrack on the target surface with a distance of 16 cm between the target surface and the orifice of the EQP. The measured energy range for the positively and negatively charged ions was from 0 to 200 eV/z (increment of 0.5 eV/z) and from 0 to 750 eV/z (increment of 1 eV/z), respectively, with z being the charge state number. In the case of the negatively charged ions, a multiplier voltage of 2600 V was used, as compared to 1800 V in the case of the positively charged ions, in order to enhance the intensity of the recorded negative ion signal. The dwell time at each data point was set to 100 ms in order to average over 10 pulses. Subsequent to the measurements, the raw data were corrected for the ion charge state by multiplying the scan voltage with z and dividing the count rate by z to account for the energy bin width. No further corrections were made to account for the energy-dependence of the acceptance angle and transfer function since they are not well known. The following ions were analysed: Nb$^{+}$, Nb$^{2+}$, O$^{+}$, O$^{2+}$, O$_{2}^{+}$, Ar$^{+}$, Ar$^{2+}$, NbO$^{+}$, NbO$^{2+}$, O$^{-}$ and O$_{2}^{-}$.

\begin{figure}
 \centering
 \includegraphics[width=8cm]{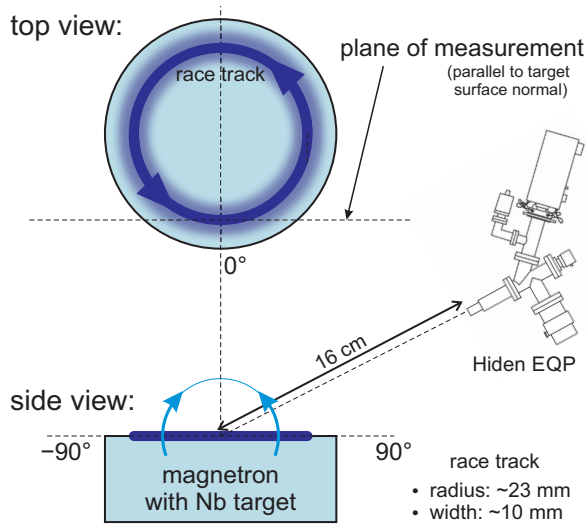}
 \caption{Schematic of the experimental setup to measure the ion energy and flux. The arrows in the race track (top view) indicate the direction of the electrons' $\mathbf{E}\times \mathbf{B}$-drift direction. The emission angle was defined in clockwise sense of rotation to 'match' the $\mathbf{E}\times \mathbf{B}$ direction.}
 \label{fig:tangential_scan_direction}
\end{figure}

In order to evaluate the influence of the ionisation zone motion on the ion flux, angular scans of the IEDFs were performed where the magnetron was placed on a rotatable table with the target surface in the pivot point as schematically shown in Fig.~\ref{fig:tangential_scan_direction}. The scans were done in tangential direction of the racetrack with an available scan range of 140\textdegree. Two individual scans were performed to cover the entire angular range. A first scan covered the angles in $+\mathbf{E}\times \mathbf{B}$ direction (ionisation zones move towards the EQP), while the $-\mathbf{E}\times \mathbf{B}$ direction (ionisation zones move away from the EQP) was measured in a second scan. The results presented in the angular range from --30 to 30\textdegree~are the average of the two scans. Angles exceeding $\pm$90\textdegree~indicate a scan direction from the backside of the magnetron, i.e.~without direct line of sight to the racetrack. As a reference also scans in radial direction of the racetrack were performed (IEDFs not shown).

\begin{figure}
 \centering
 \includegraphics[width=8cm]{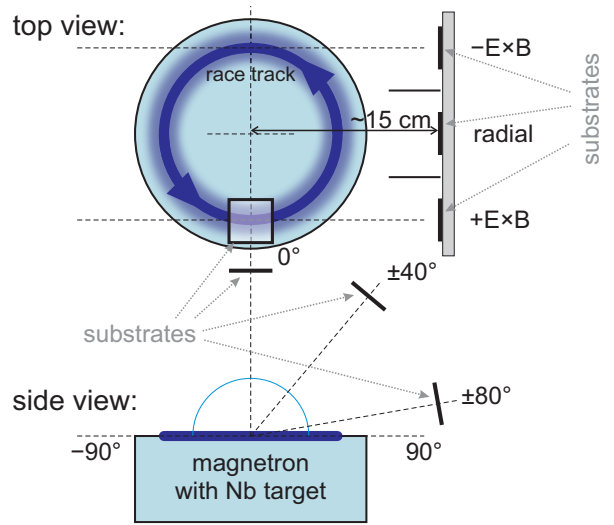}
 \caption{Schematic of the magnetron and substrate holder arrangement used for the deposition of the NbO$_x$ thin films.}
 \label{fig:depositions}
\end{figure}

Subsequent to the measurements of the IEDFs, NbO$_x$ thin films were deposited at different angles with respect to the target surface. As schematically shown in Fig.~\ref{fig:depositions}, the substrates were mounted on a holder at a distance of $\sim$15 cm to the target surface. For the depositions at 40\textdegree~and 80\textdegree, three samples were placed at different positions with respect to the racetrack and, hence, the motion of the ionisation zones: $+\mathbf{E}\times \mathbf{B}$, radial and $-\mathbf{E}\times \mathbf{B}$ direction. Due to the radius of the racetrack of $\sim$23 mm, the deposition distance slightly varied for the different substrate positions and deposition angles. Shields with a length of $\sim$3~cm were mounted between the three substrates to improve the directionality of the depositions. At a deposition angle of 0\textdegree, the substrates were mounted facing the racetrack. These samples served as references within this study. Si (about 20 mm $\times$ 10 mm $\times$ 0.5 mm), sodalime glass (25 mm $\times$ 12 mm $\times$ 1 mm) and glassy carbon (10 mm $\times$ 10 mm $\times$ 1 mm) were used as substrate materials. The deposition time was 60 min. No additional heating was applied during the depositions and the substrate holder was grounded.

The films deposited on the glassy carbon substrates were analysed by Rutherford backscattering spectrometry (RBS) in order to measure the chemical composition. The measurements were carried out using a model 5SDH Pelletron tandem accelerator manufactured by National Electrostatics Corporation generating energetic alpha beams of up to 5 MeV with terminal voltage of up to 1.7 MV. This technique detects the energies and amount of backscattered He ions from a solid target and can be used to investigate the depth profile of individual elements in a thin film (detection limit $\sim$0.5\%). The measurements were done with He$^{2+}$ of energy of 3.04 MeV, which is the energy for oxygen nuclear reaction analysis (NRA), i.e.~the sensitivity of the measurement is enhanced for oxygen due to the non-Rutherford (inelastic) cross section enhancement at that energy \cite{Nastasi2014}. Fittings of the measured RBS spectra were done using SIMNRA. The composition of the reference sample synthesised at 0\textdegree~was Nb$_{2.16\pm0.05}$O$_{4.84\pm0.05}$ were the error was estimated by considering the range of concentrations that yield acceptable fits. Possible structural changes of the NbO$_x$ films were analysed using X-ray diffractometry. The used device was a Bruker D8 Discover X-ray diffractometer operating with Cu-K$_{\alpha}$ radiation ($\lambda_1$ = 0.15406 nm and $\lambda_2$ = 0.15444 nm) and equipped with a General Area Detector Diffraction System.

For the evaluation of the film thickness with varying deposition angle, two sets of data were available. The film thickness on the glass substrates was measured at an edge using a profilometer (Veeco Dektak II). The edge was created by marking an area in the centre of the substrate with a pen prior to deposition and removal of the low adhesive film in this zone in an ultrasonic bath after deposition. In addition, the RBS data was used to calculate the film thickness assuming the Nb$_{2}$O$_{5}$ bulk density of 4.6 g/cm$^3$ \cite{Lide2004}. The molar mass was calculated according to the measured chemical composition of each film. In a similar way, the film density was calculated by dividing the RBS atomic density by the profilometer film thickness and multiplying the result by the atomic mass (derived from the molar mass).

A brief analysis of the optical properties of the deposited NbO$_x$ thin films was done by measuring the transmittance and reflectance using a PerkinElmer Lambda 950 spectrophotometer.

\section{Results}

\subsection{Ion energy and flux of the positively charged ions}

\begin{figure*}
 \centering
 \includegraphics[width=12cm]{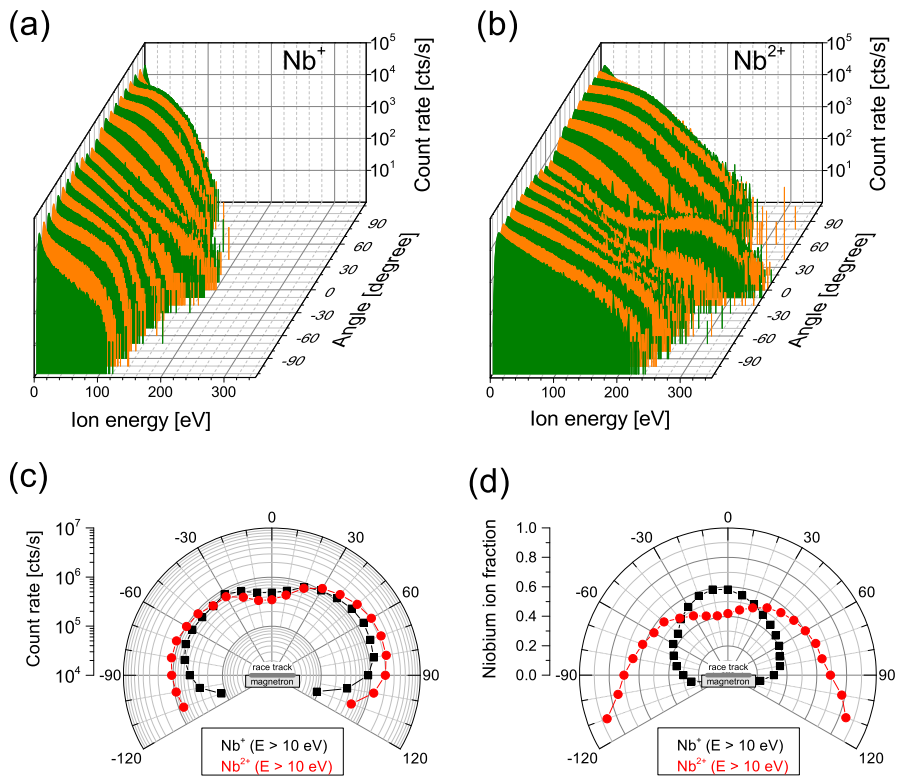}
 \caption{Angularly resolved IEDFs of (a) Nb$^+$ and (b) Nb$^{2+}$. (c) Count rates and (d) ion fractions vs.~emission angle of niobium ions. The ion count rates were calculated by integrating the niobium IEDFs for energies greater than 10 eV.}
 \label{fig:Nb-ions_IEDFs_tangential}
\end{figure*}

As shown in Fig.~\ref{fig:Nb-ions_IEDFs_tangential}, Nb$^{+}$ and Nb$^{2+}$ ions were detected in the studied HiPIMS plasma with maxima in energy of 200 and 300 eV, respectively. These maxima were recorded at an emission angle of 0\textdegree, where the formation of an extended high energy tail in the IEDFs around this angle is more pronounced for Nb$^{2+}$ as compared to the IEDFs of Nb$^{+}$. In the medium energy range at about 50 eV in the case of Nb$^{+}$ and at about 100 eV in the case of Nb$^{2+}$, there were higher count rates in the IEDFs in the angular range between 0 and 90\textdegree~than in the respective IEDFs in the range from 0 to --90\textdegree, even though the experimental setup was geometrically symmetric.

The Nb ion count rate as shown in Fig.~\ref{fig:Nb-ions_IEDFs_tangential}(c) was calculated by integrating the individual IEDFs for the different emission angles. The integration was limited to ion energies greater than 10 eV to exclude thermalised ions with random motion. In general, the integrated ion count rates of Nb$^{+}$ and Nb$^{2+}$ are similar and an increased flux sidewards (high angles with respect to target surface normal) can be noticed. However, the flux is not symmetric as more ions are emitted between 0 and 90\textdegree~as compared to the range between 0 and --90\textdegree. Fig.~\ref{fig:Nb-ions_IEDFs_tangential}(d) shows the Nb ion fractions based on the integrated IEDFs where a preferred flux of Nb$^{2+}$ sidewards with a slight asymmetry in favour of the flux between 0 and 90\textdegree~can be observed.

\begin{figure*}
 \centering
 \includegraphics[width=12cm]{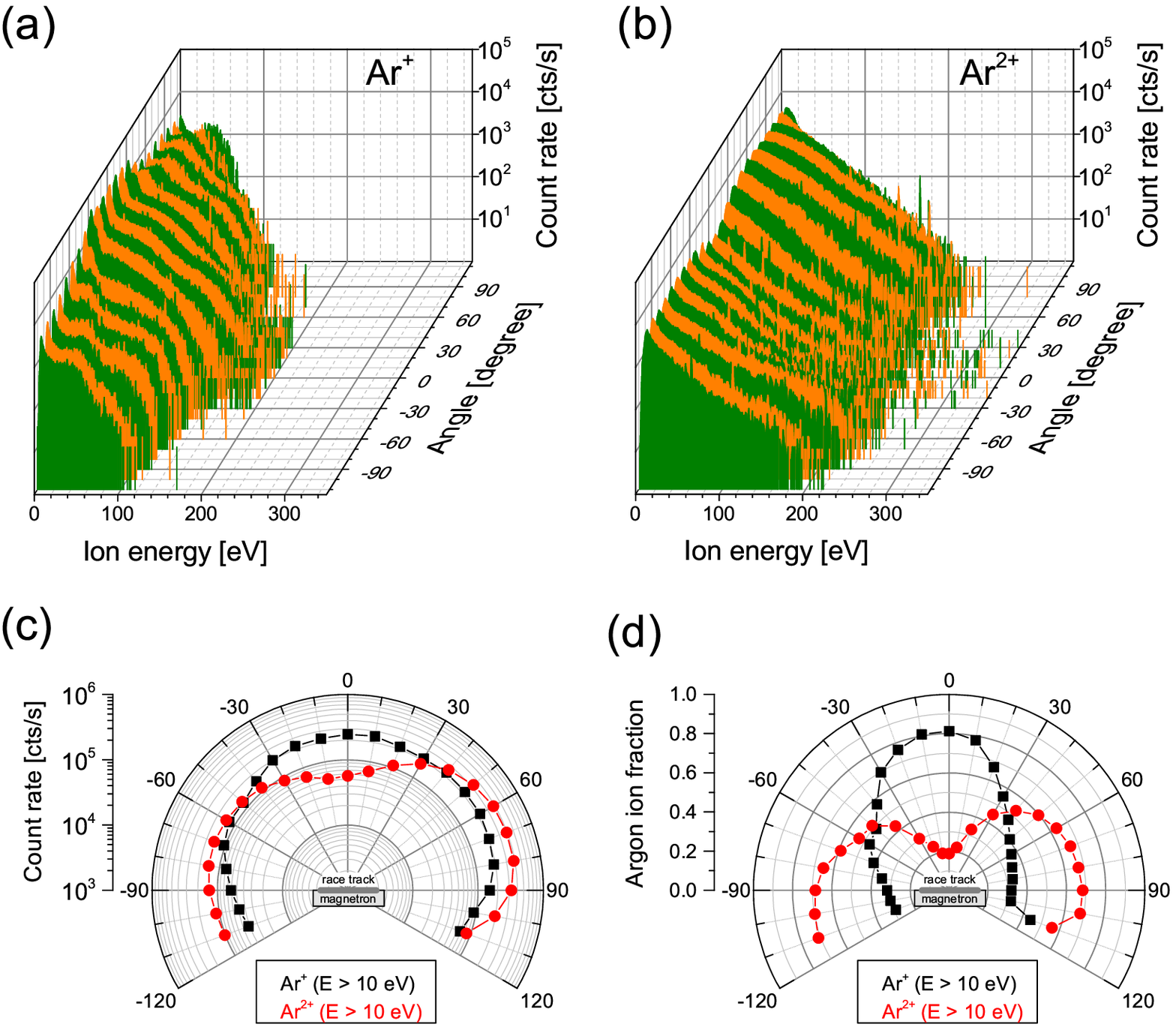}
 \caption{Angularly resolved IEDFs of (a) Ar$^+$ and (b) Ar$^{2+}$. (c) Count rates and (d) ion fractions vs.~emission angle of argon ions. The ion count rates were calculated by integrating the argon IEDFs for energies greater than 10 eV.}
 \label{fig:Ar-ions_IEDFs_tangential}
\end{figure*}

The characteristics of the Ar ions are similar to the Nb ions as shown in Fig.~\ref{fig:Ar-ions_IEDFs_tangential}. They vary in the lower maximum energy in the IEDFs of Ar$^{+}$ and Ar$^{2+}$ and in the by about one order of magnitude lower ion count rates. A further variation is the more pronounced difference in the emission angle between Ar$^{+}$ and Ar$^{2+}$ (see Fig.~\ref{fig:Ar-ions_IEDFs_tangential}(d)) as compared to the emission angles of Nb$^{+}$ and Nb$^{2+}$ (see Fig.~\ref{fig:Nb-ions_IEDFs_tangential}(d)).

\begin{figure*}
 \centering
 \includegraphics[width=16cm]{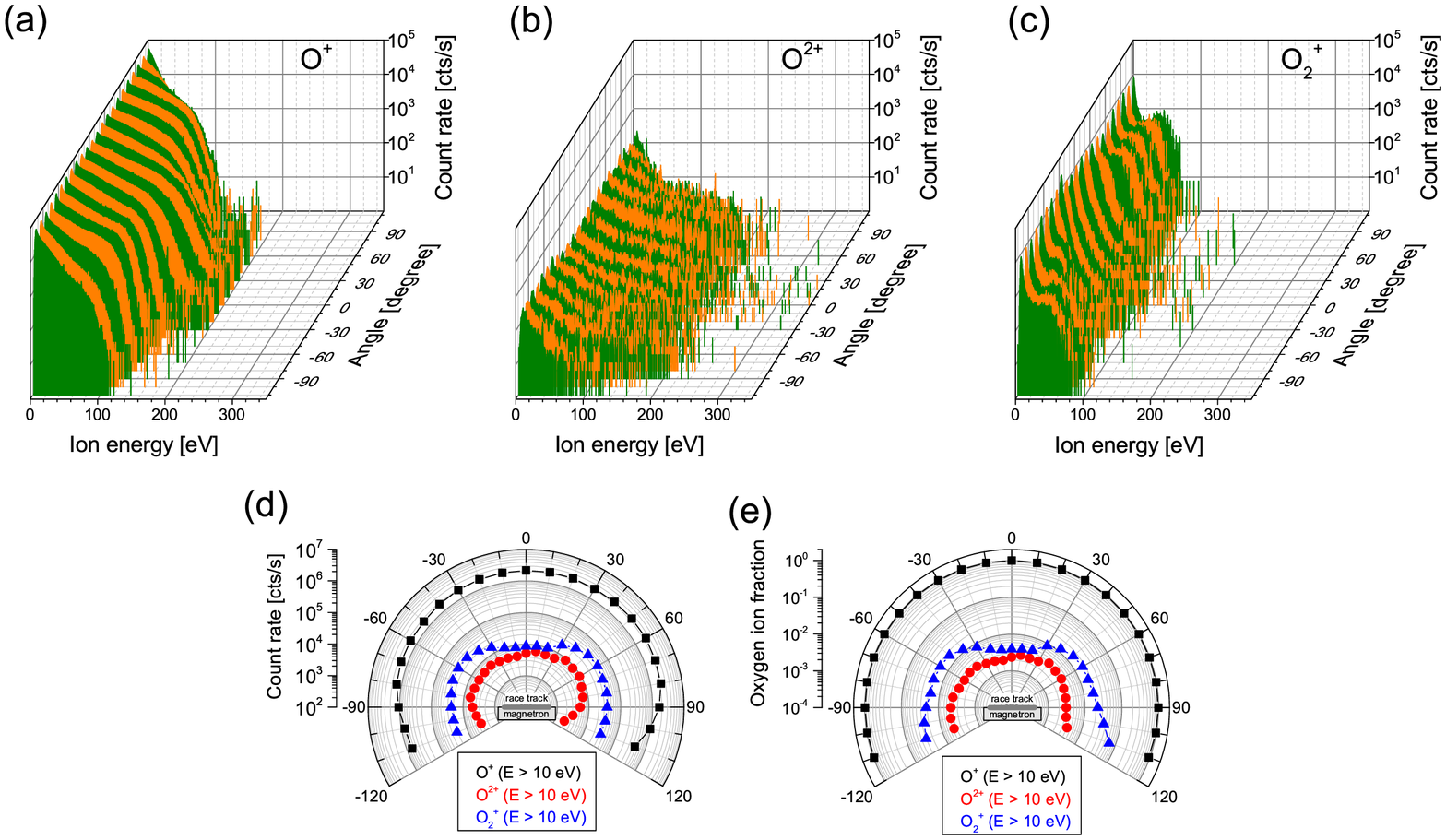}
 \caption{Angularly resolved IEDFs of (a) O$^+$, (b) O$^{2+}$ and (c) O$_2^+$. (d) Count rates and (e) ion fractions vs.~emission angle of positively charged oxygen ions. The ion count rates were calculated by integrating the oxygen IEDFs (positively charged) for energies greater than 10 eV.}
 \label{fig:O(+)-ions_IEDFs_tangential}
\end{figure*}

Among the positively charged oxygen ions, O$^+$ is the most abundant ion species (see Fig.~\ref{fig:O(+)-ions_IEDFs_tangential}). The atomic oxygen IEDFs revealed high-energy tails extending up to 200 eV, whereas the O$_2^+$ ions are mainly thermalised with an energy of a few eV. However, the angular flux of all these ions is rather isotropic as the differences in the ion count rates between positive and negative angles are smaller than in the case of the other positively charged ions.

\begin{figure*}
 \centering
 \includegraphics[width=12cm]{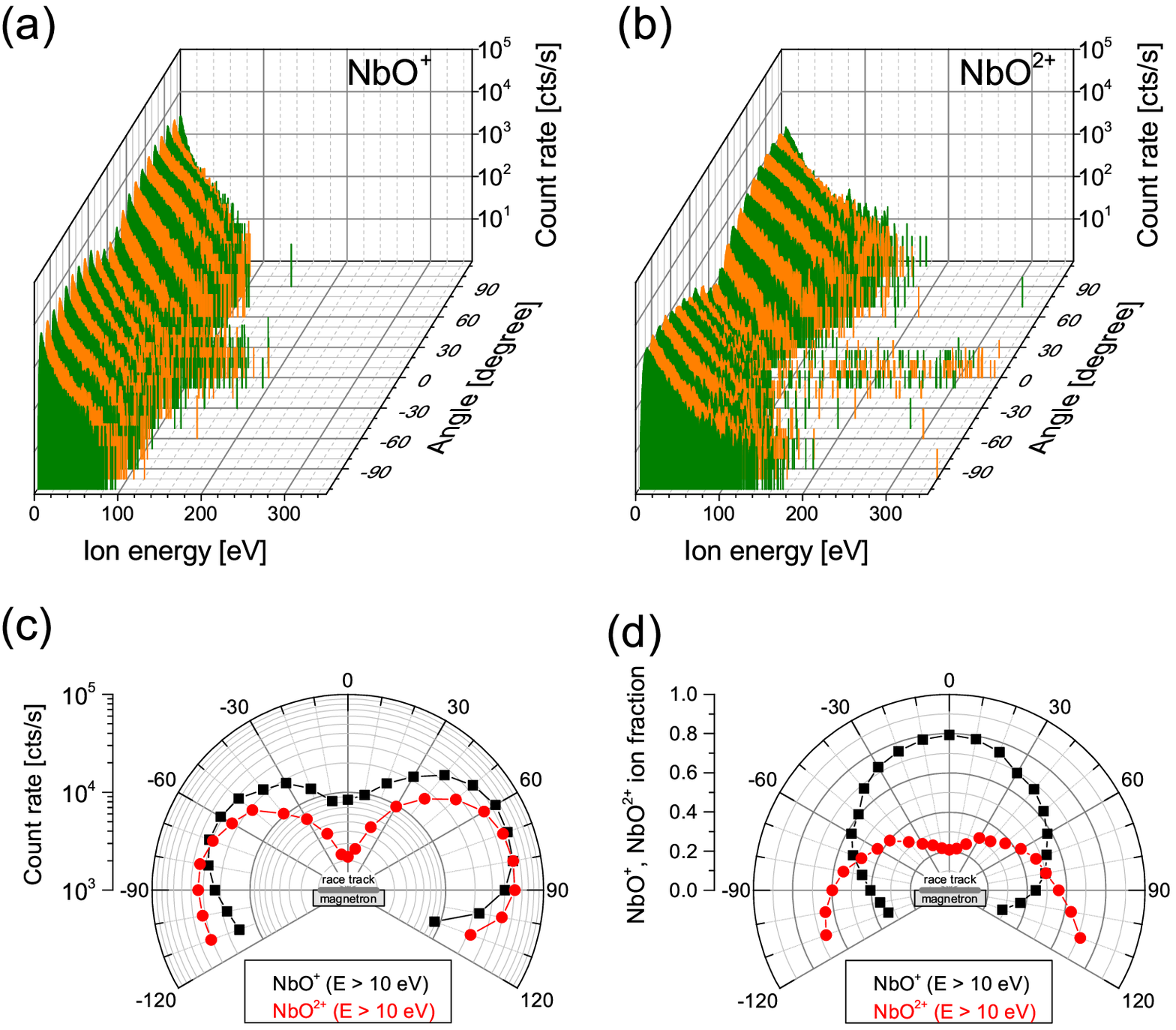}
 \caption{Angularly resolved IEDFs of (a) NbO$^+$ and (b) NbO$^{2+}$. (c) Count rates and (d) ion fractions vs.~emission angle of NbO$^+$ and NbO$^{2+}$. The ion count rates were calculated by integrating the NbO$^+$ and NbO$^{2+}$ IEDFs for energies greater than 10 eV.}
 \label{fig:NbO-ions_IEDFs_tangential}
\end{figure*}

In addition to the single-element ions described so far, also molecular NbO ions were detected as shown in Fig.~\ref{fig:NbO-ions_IEDFs_tangential}. The IEDFs of NbO$^+$ and NbO$^{2+}$ extend to maximum energies of about 100 eV while both ion species preferentially flow sidewards. Similar to the Nb ions, more ions are recorded at positive angles and the fraction of doubly charged NbO ions is higher at high angles with respect to the target surface normal. The integrated count rates of the NbO ions, however, are about two orders of magnitude lower than the integrated count rates of the Nb ions.

\subsection{Ion energy and flux of the negatively charged ions}

\begin{figure*}
 \centering
 \includegraphics[width=12cm]{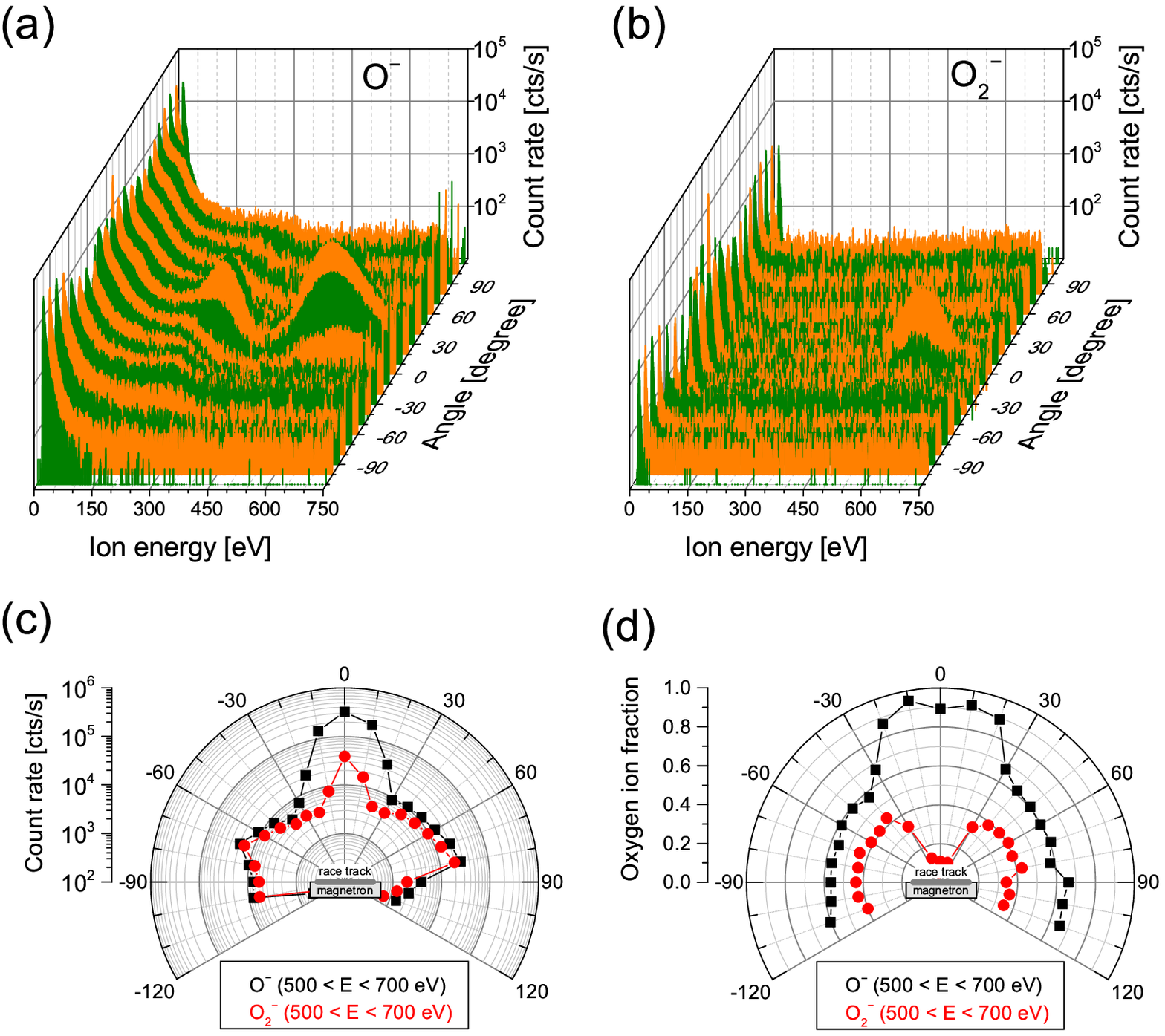}
 \caption{Angularly resolved IEDFs of (a) O$^-$ and (b) O$_2^-$. (c) Count rates and (d) ion fractions vs.~emission angle of negatively charged oxygen ions. The ion count rates were calculated by integrating the oxygen IEDFs (negatively charged) for energies between 500 and 700 eV.}
 \label{fig:O(-)-ions_IEDFs_tangential}
\end{figure*}

Due to the high electronegativity of oxygen, it is typical to encounter negatively charged oxygen ions in gas discharges with a fraction of O$_2$ in the atmosphere. As shown in Fig.~\ref{fig:O(-)-ions_IEDFs_tangential}, O$^-$ and O$_2^-$ ions were observed within the current work. Their IEDFs show three distinct energy regions: thermalised and low-energy ions with energies below 100 eV, a medium energy peak at about 300 eV (only in the case of O$^-$) and a high energy peak at about 600 eV. The latter two features are only present at angles between --20 and 20\textdegree, i.e.~at emission angles close to the target surface normal. This is corroborated by Fig.~\ref{fig:O(-)-ions_IEDFs_tangential}(c) where the ion count rates integrated over an energy range from 500 to 700 eV and, hence, the angular flux of the high-energy O$^-$ and O$_2^-$ ions is displayed. The elevated constant background signal recorded for all studied emission angles is most likely due to electrons reaching the detector and not related to negatively charged ions. However, comparing both negatively charged oxygen ions, it is apparent that O$^-$ ions represent the majority as shown by the angular ion fractions in Fig.~\ref{fig:O(-)-ions_IEDFs_tangential}(d). 

\subsection{Structure and properties of the deposited NbO$_x$ thin films}

\begin{figure}
 \centering
 \includegraphics[width=8cm]{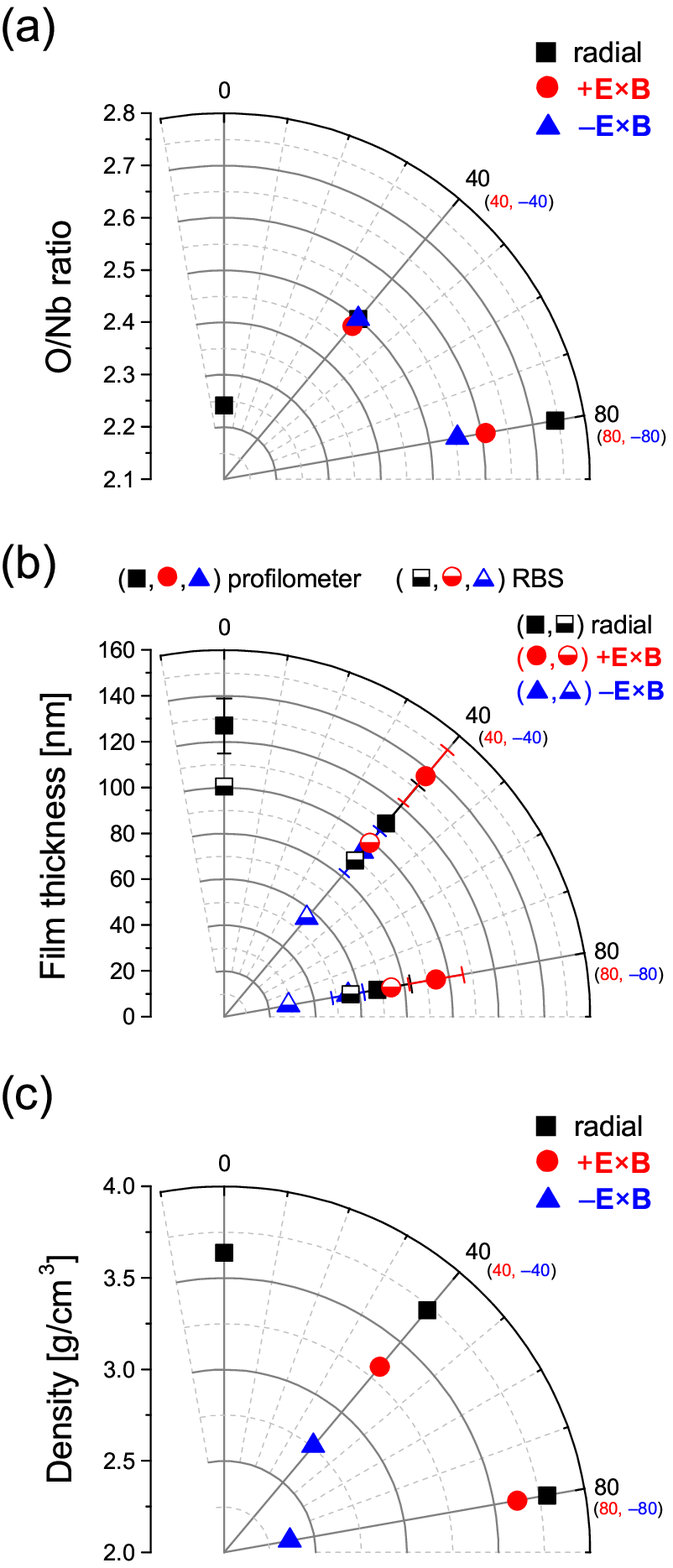}
 \caption{Selected film properties in dependence on deposition angle and direction: (a) O/Nb ratio, (b) film thickness and (c) film density.}
 \label{fig:film_properties}
\end{figure}

With an O/Nb ratio of $\sim$2.24 as shown in Fig.~\ref{fig:film_properties}(a), the reference film deposited at an angle of 0\textdegree~is slightly overstoichiometric in niobium (the O/Nb ratio of Nb$_2$O$_5$ is 2.5 and is used as a reference). With increasing deposition angle, however, the oxygen content in the films increases regardless of the position on the substrate holder. At an angle of 40\textdegree, all deposited NbO$_x$ films were perfectly stoichiometric with respect to Nb$_2$O$_5$, whereas at 80\textdegree~the films were overstoichiometric in oxygen. In the latter case, a large scattering of the O/Nb ratio was present, without apparent correlation to the deposition position.

All films revealed an amorphous structure, most likely due to the growth conditions without external heating and bias voltage.

As can be seen in Fig.~\ref{fig:film_properties}(b), the films deposited in radial direction of the racetrack show the expected reduction in film thickness with increasing deposition angle. Regarding the influence of the ionisation zone motion, it is apparent that a higher film thickness was obtained if the substrate was placed in $+\mathbf{E}\times \mathbf{B}$ than in $-\mathbf{E}\times \mathbf{B}$ direction. Even at an angle of 40\textdegree~the deposition rate in $+\mathbf{E}\times \mathbf{B}$ direction was equal or even slightly increased as compared to 0\textdegree. At 80\textdegree, the deposition rate in this direction was still about 50\%. The generally lower values of the film thickness calculated from the RBS data is most likely due to the lower thin film density than the assumed Nb$_{2}$O$_{5}$ bulk density.

Apparently, the evolution of the film thickness is influenced by a variation of the density. Fig.~\ref{fig:film_properties}(c) shows the film densities of the deposited NbO$_x$ films. As compared to the Nb$_{2}$O$_{5}$ bulk density of 4.6 g/cm$^3$ \cite{Lide2004}, all films show a reduced density. In radial direction, the film density is independent of the deposition angle. In $+\mathbf{E}\times \mathbf{B}$ direction the density is slightly reduced but remains rather constant, whereas in $-\mathbf{E}\times \mathbf{B}$ direction a strong reduction in density with increasing angle can be noticed. 

\begin{figure}
 \centering
 \includegraphics[width=8cm]{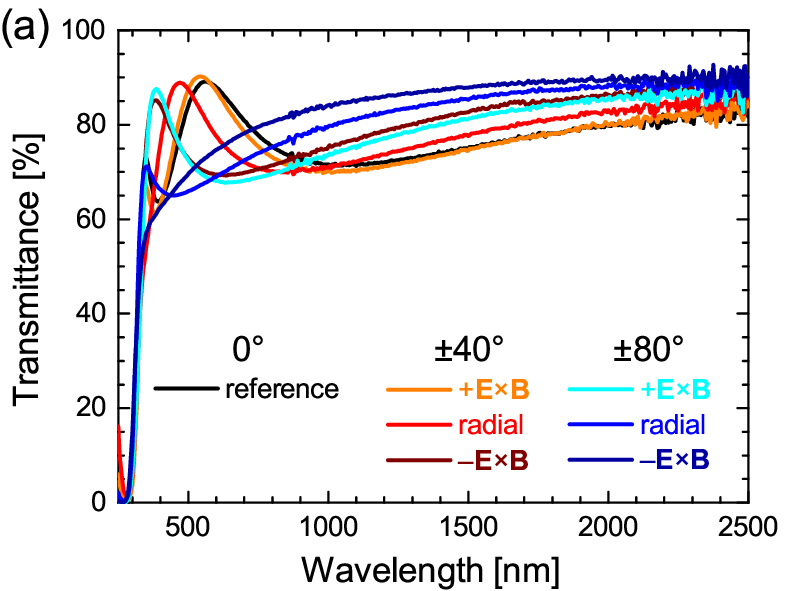}
 \includegraphics[width=8cm]{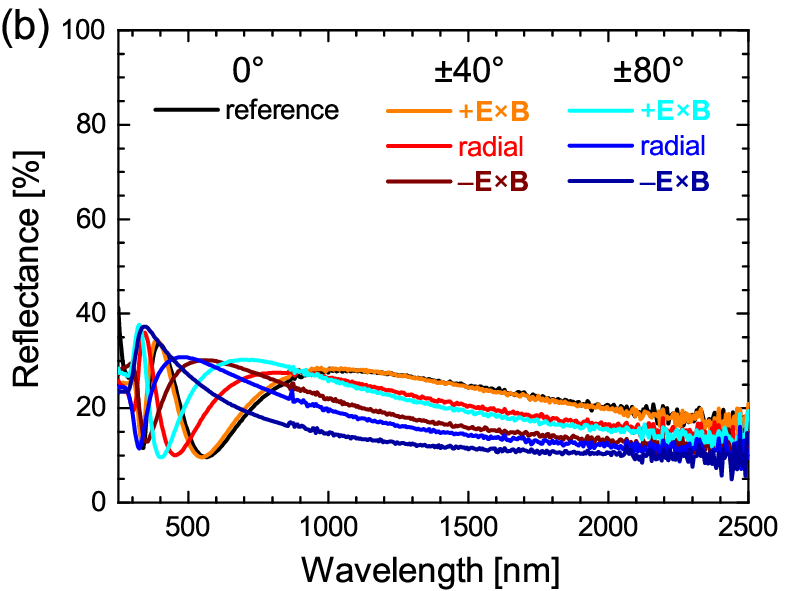}
 \caption{(a) transmittance and (b) reflectance of the deposited NbO$_x$ thin films.}
 \label{fig:optical_properties}
\end{figure}

Since NbO$_x$ or Nb$_2$O$_5$ films are frequently used as optical thin films due to their high transmittance, high refractive index and chemical stability \cite{Venkataraj2001,Lee2002,Lai2005,Cetinorgu-Goldenberg2012}, we performed a brief characterisation of their optical properties. As shown in Fig.~\ref{fig:optical_properties}, the transmittance and the reflectance in the studied wavelength range was at 70--90\% and 10--30\%, respectively, regardless of the substrate position during deposition. Variations in transmittance and reflectance at small wavelengths due to the variation in the film thickness are apparent. The absorptance is negligible throughout the entire wavelength range. The sharp cut-off at about 300 nm is due to the used glass substrate.

\section{Discussion}

\subsection{Angular ion flux and ion energy}

An asymmetry in the plasma properties of HiPIMS discharges, in particular in the ion energy, has been reported by several researchers \cite{Lundin2008,Poolcharuansin2012,Panjan2014}. The observed differences in the IEDFs between the $\mathbf{E}\times \mathbf{B}$ directions were explained by the motion of the ionisation zones and the associated potential humps \cite{Anders2013,Panjan2014}. Details about the formation of potential humps can be found in Refs. \cite{Anders2012d,Anders2013,Brenning2013,Panjan2014,Anders2014a}; here we will limit the discussion to the influence of the potential hump on the ion energy.

Ionisation zones represent perturbations of the plasma cloud in the vicinity of the target surface and are characterised by high plasma density and increased light emission as compared to other regions. They are enveloped by a double layer with ions concentrated in the centre and electrons at the outer margins, where the latter can leave the zone due to their higher mobility. The resulting electric field causes an additional $\mathbf{E}\times \mathbf{B}$ drift and the path of the drifting electrons is most likely diverted as described in \cite{Anders2014a}. However, ions, which are not magnetised in typically used magnetron sputter deposition systems, can experience an acceleration in the local electric field if they originate from the centre region of the ionisation zone. Passing through the associated potential hump, the direction of ions' motion is altered, usually off-axis with respect to the target surface normal. This effect can explain the observed broader angular ion flux in HiPIMS discharges as compared to the typical ion flux in dcMS \cite{Leroy2011,Cada2013,Panjan2014}.

In addition, the motion of the ionisation zones and the associated potential hump causes another acceleration of the ions. Ions leaving the potential hump region in the same direction as its motion can ``surf'' on the front of this plasma wave and gain an additional energy of a few 10 eV \cite{Anders2013,Panjan2014}. This effect can be noticed in the IEDFs of the positively charged ions, in particular in the IEDFs of Nb$^+$ and Nb$^{2+}$ where there is an additional contribution for the ions recorded at positive angles, i.e.~in $+\mathbf{E}\times \mathbf{B}$ direction, as compared to the ones measured at negative angles ($-\mathbf{E}\times \mathbf{B}$ direction). Even though this energy peak is rather broad, the fact, that the Nb$^{2+}$ ions gain approximately twice the energy that the Nb$^+$ ions win, points towards an acceleration mechanism in an electric field which is in agreement with the above discussed presence of potential humps in the ionisation zones.

Besides this angular asymmetry in the IEDFs, the obtained absolute values of the ion energies with maxima of up to 100 and 200 eV for the singly and doubly charged ions, respectively, are high and enable energetic growth conditions for dense thin films. Due to the high degree of ionisation in HiPIMS, ions make up a major fraction of the film forming species which is in contrast to conventional magnetron sputter deposition processes where mainly neutral atoms condense on the substrate. In addition to the kinetic energy of the ions, they also provide potential energy upon the capture of electrons when they are incorporated into the growing film \cite{Anders2014}. This energy contribution is identical to the ionisation energy and, hence, the presence of a significant amount of doubly charged ions, as it is the case in the present work, enhances this effect.

Another important aspect for the growth of thin films is the angular distribution of the ion flux. With the exception of the positively charged oxygen ions, all ion species showed an anisotropic distribution where the flux along the target surface normal was typically lower as compared to the flux sidewards. In addition, there is a difference in the flux with respect to the $\mathbf{E}\times \mathbf{B}$ direction. Higher ion fluxes were recorded in the $+\mathbf{E}\times \mathbf{B}$ direction as compared to the $-\mathbf{E}\times \mathbf{B}$ direction for the Nb, Ar and NbO ions. However, a major impact on the energetic growth conditions should result from the fact that the doubly charged Nb, Ar and NbO ions are preferably emitted sidewards, hence providing more energy for the film growth at these angles.

Comparing the integrated count rates of all positively charged ion species (see Figs.~\ref{fig:Nb-ions_IEDFs_tangential}(c)--\ref{fig:NbO-ions_IEDFs_tangential}(c)), the most abundant ion species is O$^+$ followed by both Nb and then both Ar ions. Aiempanakit \textit{et al.}~also reported the highest intensity of O$^+$ for a Ti target operated in ``poisoned'' mode \cite{Aiempanakit2013}. The plasma from an Al target operated with similar discharge parameters, however, showed the highest intensities for Al$^+$ during and for Ar$^+$ after the HiPIMS pulse. Beside this material dependence of the plasma composition in HiPIMS also the geometry of the device, in particular the direction of the gas inlet, is of importance. For the deposition of stoichiometric ZrO$_2$ thin films, Vl\v{c}ek \textit{et al.}~observed significant contents of O$^+$ and O$_2^+$ ions when the O$_2$ gas inlet pointed towards the target \cite{Vlcek2015}. Injecting the O$_2$ gas in direction towards the substrate resulted in a very Ar$^+$-rich plasma.

\subsection{Negatively charged ions}

In addition to the positively charged ions, the appearance of negatively charged ions in the sputter plasmas containing O$_2$ has been reported by several researchers \cite{Zeuner1996,Mraz2006a,Mahieu2007}. The interest in studying these ions was triggered by the fact that a fraction of the negatively charged ions revealed an energy of several 100 eV. In the current work, two ion species, O$^-$ and O$_2^-$, with peaks in their IEDFs at about 600 eV were noticed. These ions are formed on or near the target surface and then accelerated in the cathode fall, i.e.~their energy gain corresponds to the discharge voltage ($\sim$600 V in this work). Small additional energy gains most likely result from sputtering effects \cite{Welzel2012}. Other peaks in the IEDFs of negatively charged oxygen containing ions were identified to stem from dissociated molecular ions \cite{Zeuner1998,Welzel2011a,Bowes2013}. This explains the appearance of the peak at $\sim$300 eV in the IEDFs of O$^-$ which result from O$_2^-$ ions that were dissociated, most likely in a collision with an electron, into a neutral O atom and an O$^-$ ion equally sharing the initial energy of the molecular ion. Other contributions from dissociated NbO$_x^-$ ions, similar to metal-oxygen ions reported in \cite{Zeuner1998,Welzel2011a,Welzel2012,Bowes2013}, were not recorded and such ions are absent in the studied discharge within the sensitivity of the measurement.

Most of the work regarding negative oxygen ions was performed using dcMS discharges rather than HiPIMS and the acceleration mechanism appears to remain unchanged when applying energetic pulses instead of dc power \cite{Bowes2013,Sarakinos2010a}. However, with the discovery of the ionisation zones, the question arises whether the associated potential hump influences the trajectory of the negatively charged oxygen ions or not. Angular and lateral scans of the ion flux in dcMS revealed that these ions mainly travel parallel to the target surface normal which can be understood by the direction of the electric field in the sheath that is perpendicular to the target surface \cite{Mahieu2009,Pokorny2011}. Changes of the target surface topography due to the progressing erosion result in slightly diverting ion trajectories \cite{Welzel2011}. As schematically shown in \cite{Anders2014a}, the electric field is strongly distorted in the vicinity of an ionisation zone possibly altering the emission angle of negatively charged ions. However, according to the angular distribution of the O$^-$ and O$_2^-$ ion fluxes in Fig.~\ref{fig:O(-)-ions_IEDFs_tangential}(c), the high-energy ions are all within $\pm$20\textdegree~of the target surface normal. In addition, there is no difference between the two $\mathbf{E}\times \mathbf{B}$ directions indicating that the disturbance of the electric field in the potential hump has little effect on the fast negatively charged oxygen ions because the height of the potential hump is much smaller compared to the potential drop in the target sheath. Nevertheless, these very energetic ions can significantly influence the thin film growth conditions when they impact on the substrate, which will be discussed in the following.

\subsection{Influence of plasma properties on film growth}

The beneficial effects of ion bombardment for the growth of thin films using physical vapour deposition in general \cite{Messier1984,Mueller1987a,Greene1993,Ohring2002,Petrov2003} and HiPIMS in particular \cite{Samuelsson2010,Anders2010d,Sarakinos2014} have been described in great detail. For the current work, we mainly limit the discussion to effects related to the angular ion fluxes. 

Using dcMS, Mráz and Schneider deposited NbO$_x$ thin films in dependence on the O$_2$ partial pressure onto grounded non-intentionally heated substrates \cite{Mraz2006b}. Close-to-stoichiometric Nb$_2$O$_{4.7}$ (O/Nb ratio of 2.35) films were obtained in the O$^+$/O$^-$ dominant mode, i.e.~when a large number of O$_2$ molecules were dissociated to atomic O in the discharge, which can be controlled by the discharge power \cite{Mraz2006b}. According to investigations by Snyders \textit{et al.}, oxygen atoms have a higher sticking coefficient than molecular O$_2$ and, therefore, their presence enables the growth of stoichiometric oxides \cite{Snyders2001}. For the current study, this suggests that the dissociation of O$_2$ in the HiPIMS discharge is more effective at high angles as higher oxygen contents were measured in the films deposited at these positions.

However, with the high degree of ionisation in HiPIMS, the condensation of ions onto the substrate makes up a major fraction of film growth. The flux of positively charged oxygen ions is rather isotropic and, therefore, most likely without influence on differences in the stoichiometry of the films. As mentioned above, the deposited films were overstoichiometric in Nb at a deposition angle of 0\textdegree, even though the flux of niobium ions was reduced in this direction as compared to the flux at higher angles. A possible explanation would be that the chemical composition is strongly influenced by neutrals which were not analysed in detail within this work. An alternative explanation would be resputtering effects due to the bombardment with energetic negatively charged oxygen ions. In this case, preferential sputtering of O atoms from the growing film upon impact of O$^-$ (and O$_2^-$) should occur due to the matching mass and, hence, optimal momentum transfer. Since the flux of these ions is limited to a narrow range near the target surface normal, the reduced oxygen content in the film deposited at 0\textdegree~could be caused by such an effect. However, more detailed studies relating the angular ion fluxes to the stoichiometry of the synthesised thin films are needed to unambiguously identify the underlying effects.

Another important beneficial effect of the increased ion bombardment when employing HiPIMS for the growth of thin films is the higher density that can be achieved. As pointed out by Samuelsson \textit{et al.}, thin films deposited by HiPIMS typically show an increased density by 5--15\% as compared to films deposited by dcMS \cite{Samuelsson2010}. The higher degree of ionisation in HiPIMS discharges, in particular of the sputtered species, results in an increased and more effective ion bombardment of the growing film and, hence, in a higher film density and improved properties like, e.g., surface roughness \cite{Samuelsson2012b}. In the current work, the film density remains constant when the deposition angle is increased in radial direction, but the film thickness decreases (see Fig.~\ref{fig:film_properties}). Such an evolution in film thickness is in agreement with reports in literature \cite{Lundin2008,Leroy2011}. The situation is different when changing the deposition angle in tangential direction of the racetrack analogue to the measurements of the IEDFs (see Figs.~\ref{fig:tangential_scan_direction} and \ref{fig:depositions}). As shown in Fig.~\ref{fig:film_properties}, in $-\mathbf{E}\times \mathbf{B}$ direction, film density and thickness, decrease with increasing deposition angle indicating disadvantageous film growth conditions. In contrast, film density and thickness remain constant or are only slightly reduced when the deposition angle is increased in $+\mathbf{E}\times \mathbf{B}$ direction. The higher flux of ions in $+\mathbf{E}\times \mathbf{B}$ as compared to $-\mathbf{E}\times \mathbf{B}$ direction provides more film forming species and, hence, a higher deposition rate can be achieved. Further, as ions make up a large or major fraction of the film forming species, the beneficial effects of increased ion bombardment come to the fore even though no additional substrate bias was applied during deposition.

Finally, the formation of an amorphous phase for all synthesised NbO$_x$ thin films is in agreement with literature. Ngaruiya \textit{et al.} reported the formation of amorphous NbO$_x$ films in deposition by dcMS without external heating and grounded substrates \cite{Ngaruiya2004}. Apparently, the additional energy input from the high degree of ionisation in the HiPIMS discharge applied within this work to grow the NbO$_x$ films is still insufficient to trigger the growth of a crystalline NbO$_x$ phase. Similar results were reported by Hála \textit{et al.} who also obtained amorphous NbO$_x$ films regardless of the applied deposition technique: dcMS, HiPIMS or modulated pulse power magnetron sputtering \cite{Hala2012a}. All films were optically transparent which is also in agreement with the films from the current study.

\section{Conclusions}

The angular asymmetry in the ion energy distributions and ion fluxes in HiPIMS discharges with respect to the racetrack is also present in reactive oxygen-containing atmospheres. The ion flux is higher in $+\mathbf{E}\times \mathbf{B}$ than in $-\mathbf{E}\times \mathbf{B}$ direction, i.e.~more ions are emitted in the same direction as the ionisation zone and associated potential hump motion. In addition, there is a pronounced asymmetry in the ion fraction with varying emission angle as doubly charged Nb, Ar and NbO ions are preferably emitted sidewards with respect to the target surface normal. These ions also show contributions in their IEDFs in the medium energy range that have about twice the energy as the respective contributions in the IEDFs of the singly charged ions indicating an acceleration by an electric field. As expected, negatively charged oxygen ions were detected as well. Their high-energy fraction with a peak at about 600 eV is practically not influenced by the potential hump as they were only recorded along the direction of the target surface normal and since the potential hump is relatively small.

The observed plasma properties were correlated with thin film growth conditions. The deposited NbO$_x$ films showed a change in stoichiometry with variation of the deposition angle. Further, higher deposition rates were noticed in $+\mathbf{E}\times \mathbf{B}$ direction which could be correlated with the higher ion fluxes in this direction. However, no significant changes of the measured optical properties were observed regardless of the deposition angle and position.

\begin{acknowledgments}
R.~Franz gratefully acknowledges the support of an Erwin Schrödinger Fellowship by the Austrian Science Fund (FWF, Project J3168-N20) which enabled his research at LBNL. Work at LBNL is supported by the U.S.~Department of Energy under Contract No.~DE-AC02-05CH11231. 
\end{acknowledgments}

\section*{References}

\begin{thebibliography}{10}
\expandafter\ifx\csname url\endcsname\relax
  \def\url#1{\texttt{#1}}\fi
\expandafter\ifx\csname urlprefix\endcsname\relax\def\urlprefix{URL }\fi
\expandafter\ifx\csname href\endcsname\relax
  \def\href#1#2{#2} \def\path#1{#1}\fi

\bibitem{Kouznetsov1999}
V.~Kouznetsov, K.~Mac{\'{a}}k, J.~M. Schneider, U.~Helmersson, I.~Petrov, Surf.
  Coat. Technol. 122~(2-3) (1999) 290--293.

\bibitem{Helmersson2006}
U.~Helmersson, M.~Lattemann, J.~Bohlmark, A.~P. Ehiasarian, J.~T. Gudmundsson,
  Thin Solid Films 513~(1-2) (2006) 1--24.

\bibitem{Sarakinos2010}
K.~Sarakinos, J.~Alami, S.~Konstantinidis, Surf. Coat. Technol. 204~(11) (2010)
  1661--1684.

\bibitem{Gudmundsson2012}
J.~T. Gudmundsson, N.~Brenning, D.~Lundin, U.~Helmersson, J. Vac. Sci. Technol.
  A 30~(3) (2012) 030801--25.

\bibitem{Britun2014}
N.~Britun, T.~Minea, S.~Konstantinidis, R.~Snyders, J. Phys. D. Appl. Phys.
  47~(22) (2014) 224001.

\bibitem{Kozyrev2011}
A.~V. Kozyrev, N.~S. Sochugov, K.~V. Oskomov, A.~N. Zakharov, A.~N. Odivanova,
  Plasma Phys. Reports 37~(7) (2011) 621--627.

\bibitem{Anders2012a}
A.~Anders, P.~Ni, A.~Rauch, J. Appl. Phys. 111~(5) (2012) 053304.

\bibitem{Ehiasarian2012}
A.~P. Ehiasarian, A.~Hecimovic, T.~de~los Arcos, R.~New, V.~{Schulz-von der
  Gathen}, M.~B{\"{o}}ke, J.~Winter, Appl. Phys. Lett. 100~(11) (2012) 114101.

\bibitem{Winter2013}
J.~Winter, A.~Hecimovic, T.~de~los Arcos, M.~B{\"{o}}ke, V.~{Schulz-von der
  Gathen}, J. Phys. D. Appl. Phys. 46~(8) (2013) 084007.

\bibitem{Anders2012d}
A.~Anders, Appl. Phys. Lett. 100~(22) (2012) 224104.

\bibitem{Yang2015}
Y.~Yang, K.~Tanaka, J.~Liu, A.~Anders, Appl. Phys. Lett. 106 (2015) 124102.

\bibitem{Poolcharuansin2015}
P.~Poolcharuansin, F.~L. Estrin, J.~W. Bradley, J. Appl. Phys. 117~(16) (2015)
  163304.

\bibitem{Ni2012}
P.~A. Ni, C.~Hornschuch, M.~Panjan, A.~Anders, Appl. Phys. Lett. 101~(22)
  (2012) 224102.

\bibitem{Gallian2013a}
S.~Gallian, W.~N.~G. Hitchon, D.~Eremin, T.~Mussenbrock, R.~P. Brinkmann,
  Plasma Sources Sci. Technol. 22~(5) (2013) 055012.
\newblock \href {http://arxiv.org/abs/1305.5354} {\path{arXiv:1305.5354}}.

\bibitem{Hecimovic2014}
A.~Hecimovic, M.~B{\"{o}}ke, J.~Winter, J. Phys. D. Appl. Phys. 47~(10) (2014)
  102003.
\newblock \href {http://arxiv.org/abs/1305.5453} {\path{arXiv:1305.5453}}.

\bibitem{Brenning2013}
N.~Brenning, D.~Lundin, T.~Minea, C.~Costin, C.~Vitelaru, J. Phys. D. Appl.
  Phys. 46~(8) (2013) 084005.

\bibitem{Anders2013}
A.~Anders, M.~Panjan, R.~Franz, J.~Andersson, P.~Ni, Appl. Phys. Lett. 103~(14)
  (2013) 144103.

\bibitem{Anders2014a}
A.~Anders, Appl. Phys. Lett. 105~(24) (2014) 244104.

\bibitem{Andersson2013}
J.~Andersson, P.~Ni, A.~Anders, Appl. Phys. Lett. 103~(5) (2013) 054104.

\bibitem{Lundin2008}
D.~Lundin, P.~Larsson, E.~Wallin, M.~Lattemann, N.~Brenning, U.~Helmersson,
  Plasma Sources Sci. Technol. 17~(3) (2008) 035021.

\bibitem{Poolcharuansin2012}
P.~Poolcharuansin, B.~Liebig, J.~W. Bradley, Plasma Sources Sci. Technol.
  21~(1) (2012) 015001.

\bibitem{Panjan2014}
M.~Panjan, R.~Franz, A.~Anders, Plasma Sources Sci. Technol. 23~(2) (2014)
  025007.

\bibitem{Yang2014}
Y.~Yang, J.~Liu, L.~Liu, A.~Anders, Appl. Phys. Lett. 105~(25) (2014) 254101.

\bibitem{Leroy2011}
W.~P. Leroy, S.~Konstantinidis, S.~Mahieu, R.~Snyders, D.~Depla, J. Phys. D.
  Appl. Phys. 44~(11) (2011) 115201.

\bibitem{Aiempanakit2013}
M.~Aiempanakit, A.~Aijaz, D.~Lundin, U.~Helmersson, T.~Kubart, J. Appl. Phys.
  113~(13) (2013) 133302.

\bibitem{Nastasi2014}
M.~Nastasi, J.~W. Mayer, Y.~Wang, {Ion Beam Analysis: Fundamentals and
  Applications}, 1st Edition, CRC Press, Boca Raton, FL, USA, 2014.

\bibitem{Lide2004}
D.~R. Lide (Ed.), {CRC Handbook of Chemistry and Physics}, 85th Edition, CRC
  Press, Boca Raton, FL, USA, 2004.

\bibitem{Venkataraj2001}
S.~Venkataraj, R.~Drese, O.~Kappertz, R.~Jayavel, M.~Wuttig, Phys. Status
  Solidi 188~(3) (2001) 1047--1058.

\bibitem{Lee2002}
C.-C. Lee, C.-L. Tien, J.-C. Hsu, Appl. Opt. 41~(10) (2002) 2043.

\bibitem{Lai2005}
F.~Lai, M.~Li, H.~Wang, H.~Hu, X.~Wang, J.~Hou, Y.~Song, Y.~Jiang, Thin Solid
  Films 488~(1-2) (2005) 314--320.

\bibitem{Cetinorgu-Goldenberg2012}
E.~{\c{C}}etin{\"{o}}rg{\"{u}}-Goldenberg, J.-E. Klemberg-Sapieha, L.~Martinu,
  Appl. Opt. 51~(27) (2012) 6498.

\bibitem{Cada2013}
M.~{\v{C}}ada, P.~Ad{\'{a}}mek, V.~Straň{\'{a}}k, {\v{S}}.~Kment,
  J.~Olejn{\'{\i}}{\v{c}}ek, Z.~Hubi{\v{c}}ka, R.~Hippler, Thin Solid Films 549
  (2013) 177--183.

\bibitem{Anders2014}
A.~Anders, Surf. Coat. Technol. 257 (2014) 308--325.

\bibitem{Vlcek2015}
J.~Vl{\v{c}}ek, J.~Rezek, J.~Hou{\v{s}}ka, T.~Koz{\'{a}}k, J.~Kohout, Vacuum
  114 (2015) 131--141.

\bibitem{Zeuner1996}
M.~Zeuner, J.~Meichsner, J.~A. Rees, J. Appl. Phys. 79~(12) (1996) 9379.

\bibitem{Mraz2006a}
S.~Mr{\'{a}}z, J.~M. Schneider, J. Appl. Phys. 100~(2) (2006) 023503.

\bibitem{Mahieu2007}
S.~Mahieu, D.~Depla, Appl. Phys. Lett. 90~(12) (2007) 121117.

\bibitem{Welzel2012}
T.~Welzel, K.~Ellmer, J. Vac. Sci. Technol. A 30~(6) (2012) 061306.

\bibitem{Zeuner1998}
M.~Zeuner, H.~Neumann, J.~Zalman, H.~Biederman, J. Appl. Phys. 83~(10) (1998)
  5083.

\bibitem{Welzel2011a}
T.~Welzel, S.~Naumov, K.~Ellmer, J. Appl. Phys. 109~(7) (2011) 073302.

\bibitem{Bowes2013}
M.~Bowes, P.~Poolcharuansin, J.~W. Bradley, J. Phys. D. Appl. Phys. 46~(4)
  (2013) 045204.

\bibitem{Sarakinos2010a}
K.~Sarakinos, D.~Music, S.~Mr{\'{a}}z, M.~to~Baben, K.~Jiang, F.~Nahif,
  A.~Braun, C.~Zilkens, S.~Konstantinidis, F.~Renaux, D.~Cossement, F.~Munnik,
  J.~M. Schneider, J. Appl. Phys. 108~(1) (2010) 014904.

\bibitem{Mahieu2009}
S.~Mahieu, W.~P. Leroy, K.~{Van Aeken}, D.~Depla, J. Appl. Phys. 106~(9) (2009)
  093302.

\bibitem{Pokorny2011}
P.~Pokorn{\'{y}}, M.~Mi{\v{s}}ina, J.~Bul{\'{\i}}ř, J.~Lan{\v{c}}ok, P.~Fitl,
  J.~Musil, M.~Novotn{\'{y}}, Plasma Process. Polym. 8~(5) (2011) 459--464.

\bibitem{Welzel2011}
T.~Welzel, K.~Ellmer, Surf. Coat. Technol. 205 (2011) S294--S298.

\bibitem{Messier1984}
R.~Messier, A.~P. Giri, R.~A. Roy, J. Vac. Sci. Technol. A 2~(2) (1984)
  500--503.

\bibitem{Mueller1987a}
K.-H. M{\"{u}}ller, Surf. Sci. 184~(1-2) (1987) L375--L382.

\bibitem{Greene1993}
J.~E. Greene, {Low-Energy Ion/Surface Interactions during Crystal Growth from
  the Vapor Phase}, in: D.~Hurle (Ed.), Handb. Cryst. Growth, 1st Edition,
  Elsevier Science Publishers, 1993, Ch.~9, pp. 640--681.

\bibitem{Ohring2002}
M.~Ohring, {Materials Science of Thin Films - Deposition and Structure}, 2nd
  Edition, Academic Press, 2002.

\bibitem{Petrov2003}
I.~Petrov, P.~B. Barna, L.~Hultman, J.~E. Greene, J. Vac. Sci. Technol. A
  21~(5) (2003) 117--128.

\bibitem{Samuelsson2010}
M.~Samuelsson, D.~Lundin, J.~Jensen, M.~a. Raadu, J.~T. Gudmundsson,
  U.~Helmersson, Surf. Coat. Technol. 205~(2) (2010) 591--596.

\bibitem{Anders2010d}
A.~Anders, Thin Solid Films 518~(15) (2010) 4087--4090.

\bibitem{Sarakinos2014}
K.~Sarakinos, D.~Magnf{\"{a}}lt, V.~Elofsson, B.~L{\"{u}}, Surf. Coat. Technol.
  257 (2014) 326--332.

\bibitem{Mraz2006b}
S.~Mr{\'{a}}z, J.~M. Schneider, Plasma Chem. Plasma Process. 26~(2) (2006)
  197--203.

\bibitem{Snyders2001}
R.~Snyders, M.~Wautelet, R.~Gouttebaron, J.~P. Dauchot, M.~Hecq, Surf. Coat.
  Technol. 142-144 (2001) 187--191.

\bibitem{Samuelsson2012b}
M.~Samuelsson, D.~Lundin, K.~Sarakinos, F.~Bj{\"{o}}refors, B.~W{\"{a}}livaara,
  H.~Ljungcrantz, U.~Helmersson, J. Vac. Sci. Technol. A 30~(3) (2012) 031507.

\bibitem{Ngaruiya2004}
J.~M. Ngaruiya, O.~Kappertz, S.~H. Mohamed, M.~Wuttig, Appl. Phys. Lett. 85~(5)
  (2004) 748.

\bibitem{Hala2012a}
M.~H{\'{a}}la, J.~{\v{C}}apek, O.~Zabeida, J.~E. Klemberg-Sapieha, L.~Martinu,
  J. Phys. D. Appl. Phys. 45~(5) (2012) 055204.

\end{thebibliography}

\end{document}